\documentclass[hidelinks, conference]{IEEEtran}
\IEEEoverridecommandlockouts{}
\usepackage{cite}
\usepackage{svg}
\usepackage{tabularray}
\usepackage{amssymb}
\usepackage{algorithmic}
\usepackage{booktabs}
\usepackage{caption}
\usepackage{graphicx}
\usepackage{textcomp}
\usepackage{lipsum}
\usepackage{pdfpages}
\usepackage{enumitem}
\usepackage{url}
\usepackage{hyperref}
\usepackage[normalem]{ulem}
\usepackage{subcaption}
\usepackage{listings}

\usepackage[most]{tcolorbox}
\colorlet{lightgray}{lightgray!40!}
\tcbset{
    on line,
    boxsep=2pt, left=0pt,right=0pt,top=0pt,bottom=0pt,
    colframe=lightgray,colback=lightgray,
    highlight math style={enhanced}
}
\usepackage{todonotes}
\usepackage{changebar}

\def\BibTeX{{\rm B\kern-.05em{\sc i\kern-.025em b}\kern-.08em
    T\kern-.1667em\lower.7ex\hbox{E}\kern-.125emX}}
\begin{document}

\title{Full-stack evaluation of Machine Learning inference workloads for RISC-V systems \vspace{-5mm}}

\author{
    \IEEEauthorblockN{Debjyoti Bhattacharjee, Anmol, Tommaso Marinelli, Karan Pathak, Peter Kourzanov}
    \IEEEauthorblockA{\em imec, Kapeldreef 75, 3001 Leuven, Belgium. ~~~ \{first\}.\{last\}@imec.be}\vspace{-42mm}
}

\maketitle

\begin{abstract}

Architectural simulators hold a vital role in RISC-V research, providing a crucial platform for workload evaluation without the need for costly physical prototypes. They serve as a dynamic environment for exploring innovative architectural concepts, enabling swift iteration and thorough analysis of performance metrics. As deep learning algorithms become increasingly pervasive, it is essential to benchmark new architectures with machine learning workloads. The diverse computational kernels used in deep learning algorithms highlight the necessity for a comprehensive compilation toolchain to map to target hardware platforms.
This study evaluates the performance of a wide array of machine learning workloads on RISC-V architectures using gem5, an open-source architectural simulator. Leveraging an open-source compilation toolchain based on Multi-Level Intermediate Representation (MLIR), the research presents benchmarking results specifically focused on deep learning inference workloads. Additionally, the study sheds light on current limitations of gem5 when simulating RISC-V architectures, offering insights for future development and refinement.
\end{abstract}
\begin{IEEEkeywords}
compilation, embedded, ML, performance \vspace{-3mm}
\end{IEEEkeywords}
\vspace{2mm}
\section{Introduction}\label{sec:intro}
The rapid advancement of deep learning~(DL) and their pervasive application across various domains have propelled the need for efficient and accurate simulation platforms to evaluate performance and optimize hardware implementations. Functional simulators such as QEMU~\cite{bellard2005qemu} and Spike~\cite{spike} can be used for building software and evaluation of correctness for RISC-V based platforms. They typically do not offer the scope for detailed performance evaluation of a workload. gem5~\cite{gem5}, on the other hand, offers the possibility to evaluate representative performance of workloads, based on a modular description of various architectural components. Furthermore, gem5 has timed models of different kinds of CPUs, such as in-order cores, out-of-order cores, etc., which helps in evaluating different hardware architectures relatively fast by changing simulation configuration parameters.

A large number of machine learning 
frameworks have evolved over time, such as PyTorch, Tensorflow, ONNX, Caffe, etc.  Even though most of the frameworks are available with a Python frontend, they have their own operator sets along with specific implementations of runtime. From the perspective of hardware developers, this leads to high development and testing overhead, since the runtime has to be ported for the specific hardware. In order to get around this problem, we focus on using MLIR representation of the ML models~\cite{lattner2021mlir}. MLIR offers a modern open source compiler infrastructure for multi-level intermediate representations and it resides as a sub-project inside LLVM~\cite{lattner2004llvm}. We use the open-source IREE ({\em Integration, Representation, and Execution Environment}) framework for compilation and runtime support. IREE has been developed by Google, for compiling, optimizing, and executing machine learning models  represented in MLIR across a variety of hardware targets~\cite{liu2022tinyiree}. IREE offers a POSIX-based runtime that relies on a lightweight virtual machine to address the challenge of deploying machine learning models across diverse hardware environments efficiently.

\begin{figure}[]
    \centering
    \begin{subfigure}[t]{0.48\columnwidth}
        \centering
        \caption{\vspace{0.1cm}\footnotesize MLIR based flow}
        \label{fig:overall_flow}
        \includegraphics[height=3.5cm]{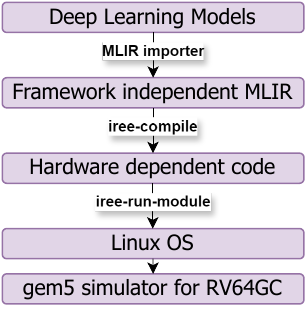}
    \end{subfigure}
    \hfill
    \begin{subfigure}[t]{0.48\columnwidth}
    \caption{\vspace{0.2cm}\footnotesize Simulated system specifications}
        \label{fig:subfig2}
    \scalebox{0.8}{\footnotesize
        \centering
        \begin{tabular}{r|l} \toprule
\textbf{Attribute}                         & \textbf{Type/version}                   \\ \midrule 
\textbf{Core Type}                         & MinorCPU, O3CPU \\
\textbf{Core Freq.}                    & 2 GHz               \\
\textbf{L1 Cache}                          & 64KB, 4-way        \\
\textbf{L2 Cache}                          & 8MB, 4-way              \\
\textbf{DRAM Type}                          & simpleMem        \\
{\textbf{DRAM Size}} & 3GB \\ 
{\textbf{DRAM Freq.}} & 1 GHz \\ 
\textbf{Kernel} & Linux v6.6.20 \\
\textbf{Bootloader} & OpenSBI v1.4 \\
\textbf{IREE Version} & 20230209.425 \\
\textbf{gem5 version} & v23.1 develop \\ 
\toprule
\end{tabular}
}    
    \end{subfigure}
    \vspace{0.2cm}
    \caption{Overall benchmarking flow and simulator configuration.}
    \label{fig:mainfigure}
\vspace{-0.6cm}
\end{figure}

\begin{table}[ht]
{\footnotesize 
\centering
\caption{List of benchmarks used for evaluation.}
\label{tab:benchmarks}
\resizebox{\columnwidth}{!}{
\begin{tabular}{rlrrrr}\bottomrule
\textbf{Task} & \textbf{Benchmark} & \textbf{Input Dimension(s)} & \textbf{Size(MB)} & \\\hline
Segmentation  & Deeplab V3 & $1\times257\times257\times3\times$float32 &2.70  & \\ \hline
Segmentation  & Densenet & $1\times224\times224\times3\times$float32 &  41.15& \\ \hline
TextDetection  & East & $1\times320\times320\times3\times$float32 &  23.03& \\ \hline
Vision  & Efficientnet lite0 & $1\times224\times224\times3\times$uint8 &5.00  & \\ \hline
LargeLanguageModel  & GPTTwo & $1\times64\times$int32 &   472.82& \\ \hline
Stylization  & Imagestylization & $1\times256\times256\times3\times$float32  &  9.00& \\ \hline
Classification  & Inception V4 & $1\times299 \times 299 \times 3 \times$float32  &  162.77& \\ \hline
CreativeAI  & Magenta & $1\times256\times256\times3\times$float32  &2.50  & \\ \hline
DepthEstimation  & Midas & $1\times256\times256\times3\times$float32  &  63.26& \\ \hline
DigitRecognition  & MNIST(Lenet5) & $1\times28\times28\times1\times$float32  &11.60  & \\ \hline
Classification & Mobilenet V1 & $1\times224\times224\times3\times$float32 &16.12  & \\ \hline
Classification & Mobilenet V2 & $1\times1001\times$float32 & 13.30 & \\ \hline
PoseEstimation  & Posenet & $1\times192\times192\times3\times$int8  &4.70  & \\ \hline
Classification  & Resnet~(50) & $1\times224\times224\times3\times$int8 &  25.09& \\ \hline
Classification  & Resnet 50 & $1\times224\times224\times3\times$float32  &  50.00 & \\ \hline
Classification  & Squeezenet & $1\times224\times224\times3\times$float32 & 4.80 & \\ \hline
ObjectDetection  & SSD Mobilenet V1 & $1\times320\times320\times3\times$uint8  & 6.65  & \\ \hline

\toprule
\end{tabular}}
}
\end{table}
\vspace{-0.3cm}
\section{Experimental Setup and Results}
While custom instructions and hardware accelerators offer performance improvements for specific machine learning tasks, we choose the RISC-V \texttt{rv64gc} architecture as a baseline
for benchmarking purposes. Firstly, employing the baseline
architecture ensures a standardized and repeatable benchmark
environment. Unlike custom architectures, which may be optimized for specific tasks, the RISC-V instruction set remains general-purpose, allowing for the evaluation of a wider range
of models and algorithms. The overall flow where we stipulate the necessity of leveraging standard tools and frameworks, portability and reuse is shown in Figure~\ref{fig:overall_flow}.

  The focus of the study is two-fold. First, it seeks to establish a foundational framework for the accurate evaluation of benchmarks' functionality. This involves the creation of a comprehensive test-bench capable of automated execution of ML benchmarks using IEEE runtime across diverse target platforms, including CPU architectures such as x86 and aarch64, GPUs like NVIDIA, emulators such as spike and QEMU, and simulators like gem5. Subsequently, extensive performance evaluation of each benchmark is conducted, assessing simulation time overheads and baselines. Throughout both aspects of the study, emphasis is placed on result repeatability and comprehensive automation across the entire stack.

\begin{figure}[ht]
    \centering
    \includegraphics[width=1.0\columnwidth]{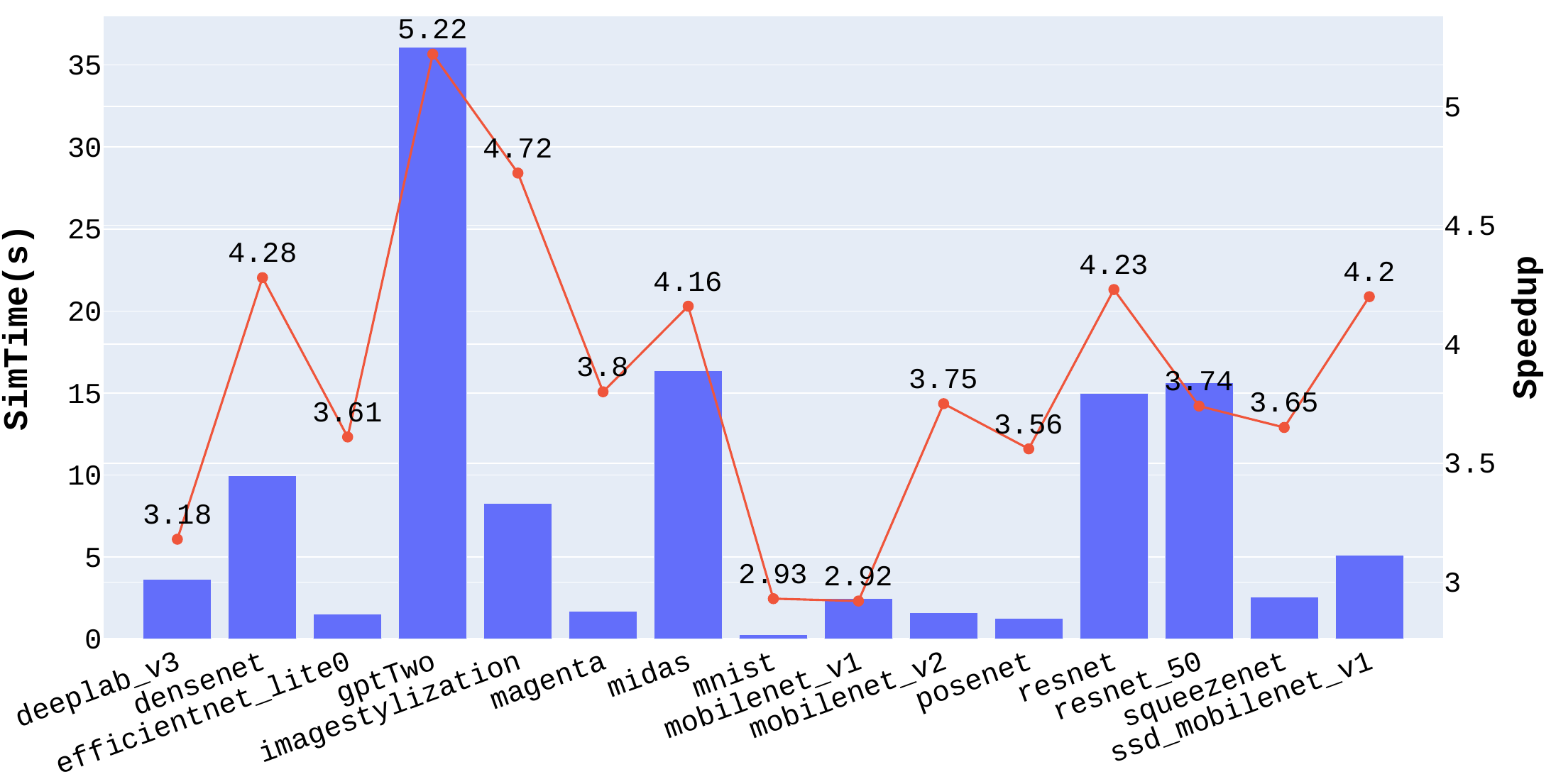}
    \caption{Performance of the workloads running on in-order CPU~(Minor CPU) and speedup on  running upon O3 CPU.}
    \label{fig:perf_eval}
    \vspace{-6mm}
\end{figure}
\begin{figure}[ht]
    \centering
    \includegraphics[width=1.0\columnwidth]{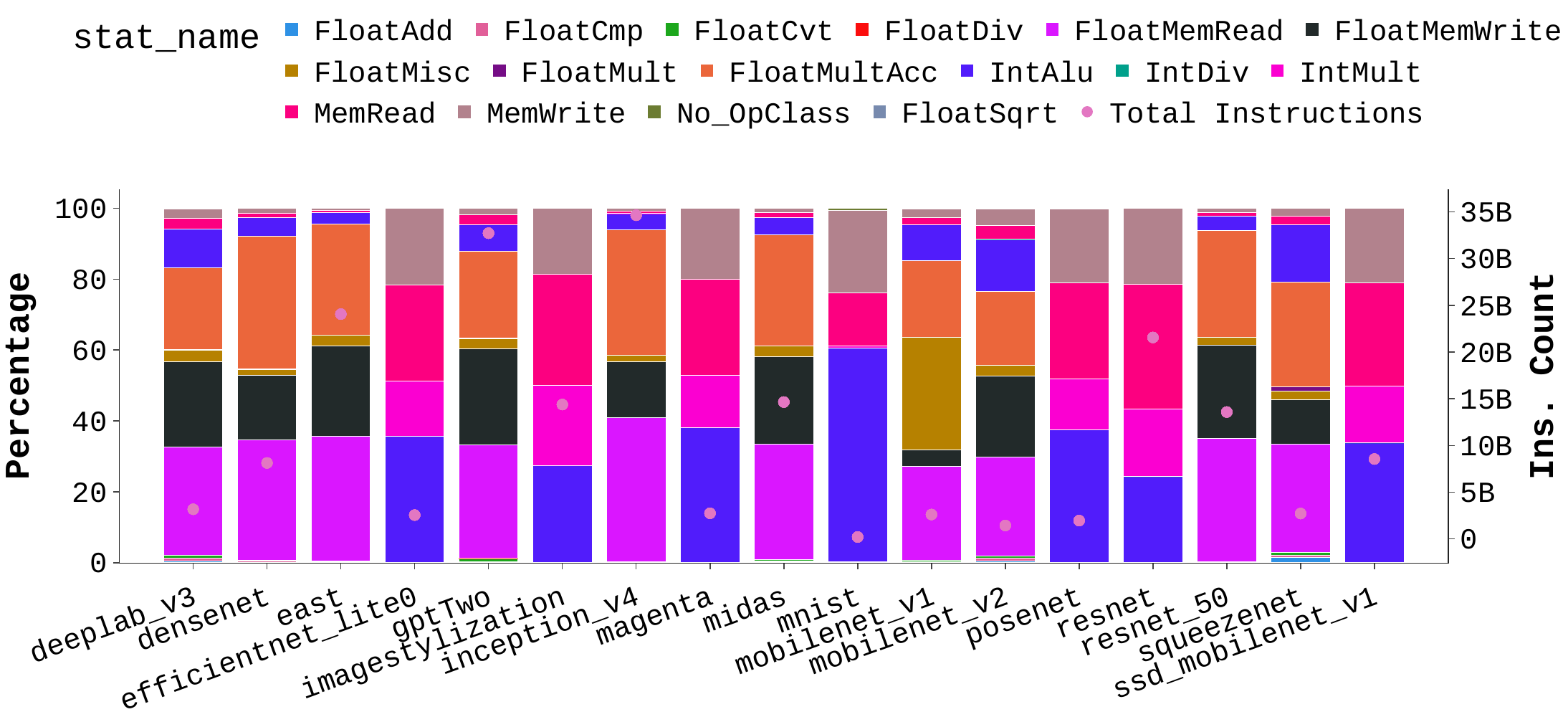}
    \caption{Instruction mix for each workload (from gem5).}
    \label{fig:ins_count}
    \vspace{-2mm}
\end{figure}

\noindent \textbf{\em Evaluation of functional correctness:}
For the considered benchmarks, the \texttt{rv64gc} results in same outputs as the host execution as well as against QEMU. This confirms the correctness of instruction implementation in the gem5 simulator. 

The early implementation of RISC-V vector instructions~\cite{rvspec} in gem5 exhibits significant issues. At the time of the experiments, our tests revealed execution bugs and discrepancies between computed and reference outputs from \texttt{rv64gv} across multiple benchmarks.
While some of the issues have been recently fixed, these findings underscore the need for refinement and optimization of gem5's RISC-V vector instruction support.

\begin{figure}[ht]
    \vspace{-4mm}
    \centering
    \includegraphics[width=0.8\columnwidth]{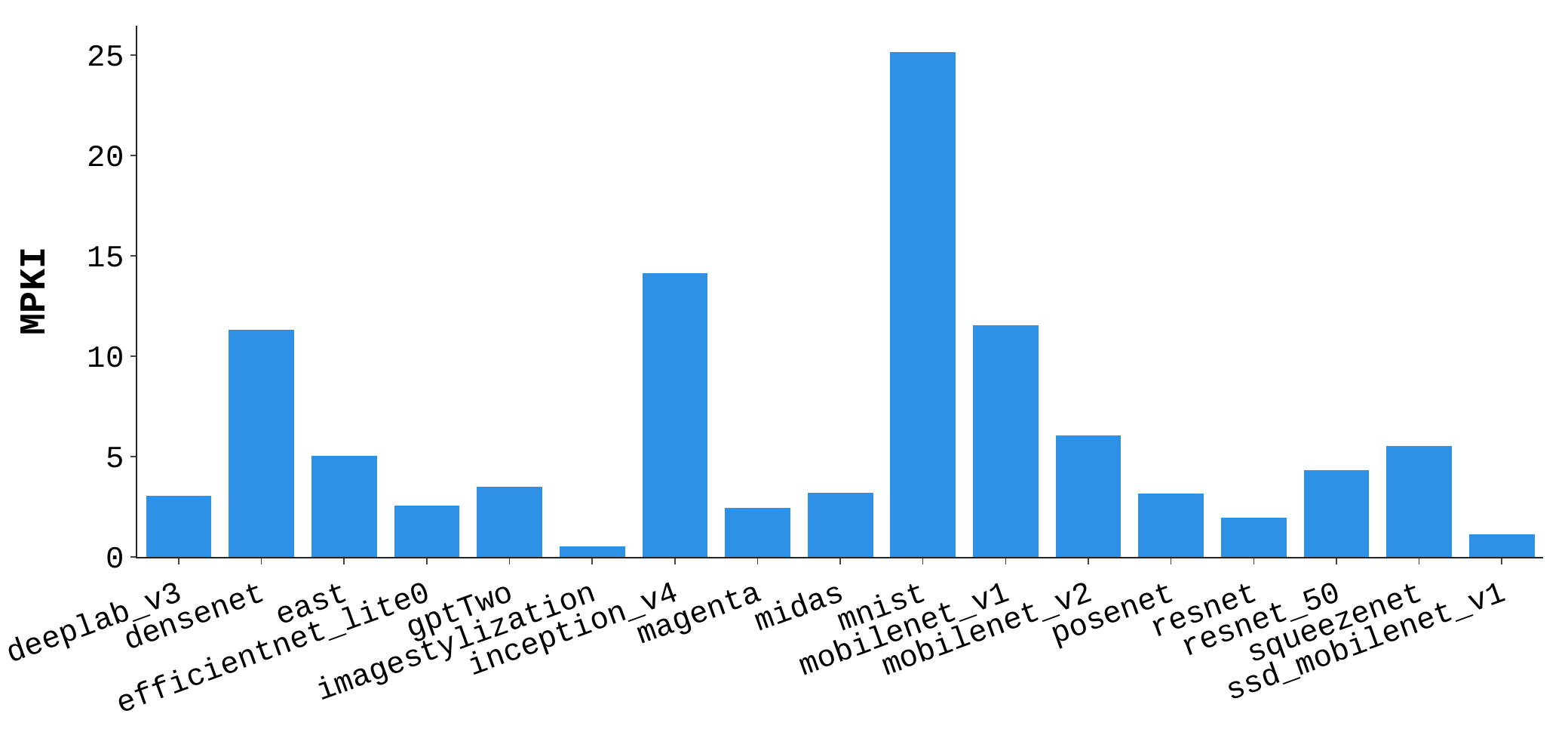}
    \caption{Miss Per Kilo Instruction~(MPKI) observed at L2 cache.}
    \label{fig:mpki}
    \vspace{-2.4mm}
\end{figure}

\noindent  \textbf{\em Performance evaluation study:} The normalised \emph{Execution Time} of the workloads in gem5 has been depicted in Figure~\ref{fig:perf_eval} for two different hardware configurations --- one with an in-order CPU~(MinorCPU) and the speedup on running it on an out-of-order~(O3) CPU model, with the same un-core configuration.  The O3 CPU offers significant performance advantage~(upto $5.22\times$) compared to the in-order CPU for all the models. Fig.~\ref{fig:ins_count} presents the breakdown of the various instructions executed by each workload. As the machine learning workloads are heavily memory dominated to due access of weights and activations, we can see a significant percentage of instructions to be memory access(MemRead and MemWrite). Also, the layers of individual neural networks use multiply accumulate operations, which show up as FloatMultAcc instructions in the instruction mix. Furthermore, we can observe the miss rate per kilo-instructions~(MPKI) in Fig.~\ref{fig:mpki}. The {\em mnist} model which is a lenet5 implementation has very few instructions, that leads to a high MPKI. For the larger benchmarks, the MPKI is sufficiently low. 

\begin{figure}[ht]
    \centering
    \vspace{-2mm}
    \includegraphics[width=0.8\columnwidth]{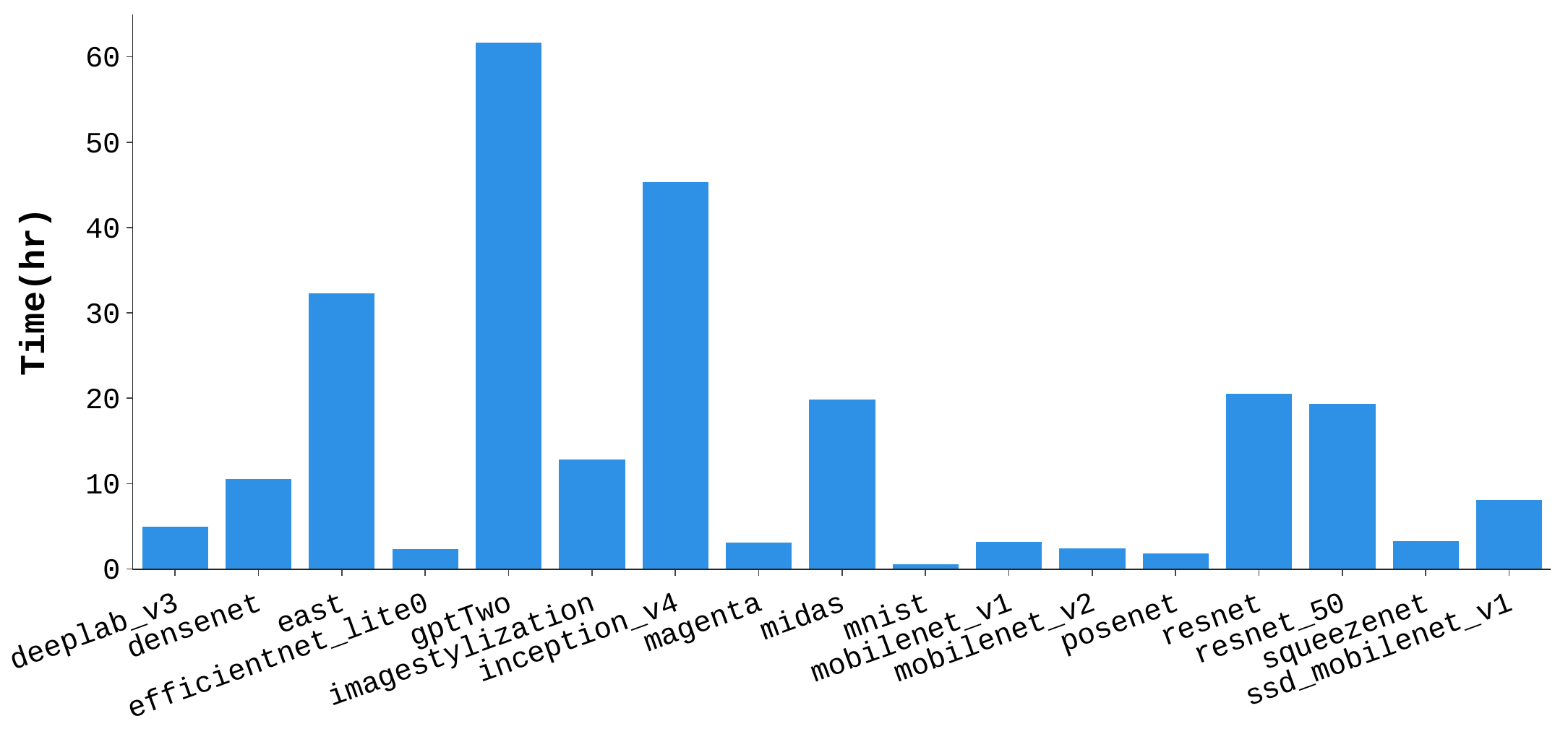}
    \caption{Time~(in hours) for simulation of workload on a x86 host.}
    \label{fig:simtime}
    \vspace{-2mm}
\end{figure}
gem5, being a single threaded architectural simulator takes an extremely large time to simulate large workloads such as {\em gptTwo}, as evident from Fig.~\ref{fig:simtime}. However, the fine modeling granularity in gem5 and support for full-system simulation outweighs the lower simulation throughput (compared to other functional simulators like QEMU and Spike). 
\section{Conclusion and Future Work}
Our current focus is on concluding the analysis and testing of additional benchmarks, such as those from MLPerf. We intend to transition from Linux to a lightweight intermediate POSIX layer, such as Zephyr. Furthermore, we aim to validate the performance of the gem5 simulator against an FPGA-based softcore implementation. These steps are integral to advancing research in the domain of RISC-V architecture simulation and development.

\vspace{-4mm}
\bibliographystyle{ieeetr}

\bibliography{ref}

\begin{thebibliography}{1}

\bibitem{bellard2005qemu}
F.~Bellard, ``{QEMU, a fast and portable dynamic translator.},'' in {\em USENIX annual technical conference, FREENIX Track}, vol.~41, pp.~10--5555, California, USA, 2005.

\bibitem{spike}
``{Spike RISC-V ISA Simulator}.'' \url{https://github.com/riscv-software-src/riscv-isa-sim}.
\newblock Accessed: 2024-03-12.

\bibitem{gem5}
J.~L. Power and et~al., ``The gem5 simulator: Version 20.0+,'' {\em CoRR}, vol.~abs/2007.03152, 2020.

\bibitem{lattner2021mlir}
C.~Lattner and et~al., ``{MLIR: Scaling compiler infrastructure for domain specific computation},'' in {\em 2021 IEEE/ACM International Symposium on Code Generation and Optimization (CGO)}, pp.~2--14, IEEE, 2021.

\bibitem{lattner2004llvm}
C.~Lattner and V.~Adve, ``{LLVM: A compilation framework for lifelong program analysis \& transformation},'' in {\em International symposium on code generation and optimization, 2004. CGO 2004.}, pp.~75--86, IEEE, 2004.

\bibitem{liu2022tinyiree}
H.-I.~C. Liu, M.~Brehler, M.~Ravishankar, N.~Vasilache, B.~Vanik, and S.~Laurenzo, ``{TinyIREE: An ML execution environment for embedded systems from compilation to deployment},'' {\em IEEE Micro}, vol.~42, no.~5, pp.~9--16, 2022.

\bibitem{rvspec}
``{Vector Extension 1.0}.'' \url{https://github.com/riscv/riscv-v-spec/releases/tag/v1.0}.
\newblock Accessed: 2024-03-12.

\end{thebibliography}

\end{document}